# Memristor – The fictional circuit element

*Isaac Abraham

DCG Silicon Development, DCG, Intel Corporation.

isaac.abraham@intel.com

The memory resistor abbreviated memristor was a harmless postulate in 1971. In the decade since 2008, a device claiming to be the missing memristor is on the prowl, seeking recognition as a fundamental circuit element, sometimes wanting electronics textbooks to be rewritten, always promising remarkable digital, analog and neuromorphic computing possibilities. A systematic discussion about the fundamental nature of the device is almost universally absent. This report investigates the assertion that the memristor is a fundamental passive circuit element, from the perspective that electrical engineering is the science of charge management. With a periodic table of fundamental elements, we demonstrate that there can only be three fundamental passive circuit elements. The ideal memristor is shown to be an unphysical active device. A vacancy transport model further reveals that a physically realizable memristor is a nonlinear composition of two resistors with active hysteresis.

## Introduction

The basic question of the "missing circuit element" is whether we can we have a new passive element that cannot be made from the combination of existing passive elements. The capacitor ($C$), resistor ($R$) and inductor ($L$) are the three fundamental passive elements that a contemporary electrical engineer is familiar with. The word fundamental means fundamental passive in the rest of this document. We start the study by generating charge-voltage relationships for $C$, $R$ and $L$. Then we attempt to integrate the memristor into the mix.

Throughout the report, we use lower case symbols for charge ($q$), voltage ($v$) etc. to accommodate the most general AC representation. Phenomenological constants are in upper case. For clarity in discussion, we assume that non idealities like leakage, temperature, noise or voltage coefficients are absent. Only the phenomenon being discussed is in effect. While units may not always be explicitly stated, we assume SI units[1].

Figure 1 represents the essence of electrical engineering (EE). In Figure 1, charge is shown as a red dot with the inset plus sign, and magnetic fields where they exist are shown as dashed, curved arrows.

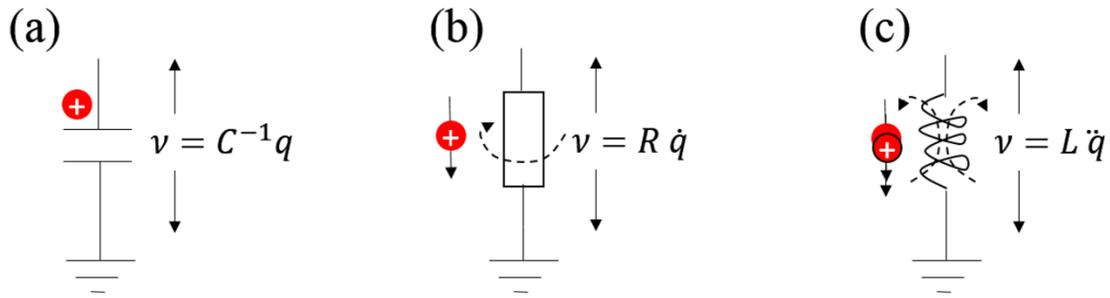

*Figure 1: Charge management underscores electrical engineering. (a) Charge on a capacitor generates voltage. (b) Charges moving at a measurable rate through a resistor generate voltage and a magnetic field; however a magnetic field will not induce a voltage in the resistor. (c) Charges moving at a rate of rate through an inductor generate a voltage and magnetic field; and a changing magnetic field will induce a voltage across the inductor.*

Known fundamental elements are passive, physically realizable devices that convert a state of rest or motion of charge into a measurable voltage across a specified element.

Table 1 introduces concepts that appear in this report and list some example devices that conform to the concepts. We follow the guidance that a regular non-Esaki pn diode is a nonlinear passive element because it does not amplify power[2].

*Table 1: Terminology.*

| # | Concept | Description | Sample devices |
|---|---------|-------------|----------------|
| 1 | Linear | Linear implies that a doubling of the input signal produces a doubling of the output signal. | $C, R, L$. |
| 2 | Nonlinear | Description 1 does not apply. | Diode, transistor. |
| 3 | Active | A physical device that can produce power gain. | Transistor. |
| 4 | Passive | A physical device that cannot produce power gain. | $C, R, L$, diode, switch. |
| 5 | Composite | A device that can be modeled from fundamental components. | $R \pm j X$, potentiometer. |
| 6 | Fundamental | Irreducible electrical representation of linear, passive elements. | $C, R, L$. |

Consider a charge that is separated from its reference plane by applying some energy. This results in a static charge placed at some location, with a potential. The potential is the work that will be done by this charge as it travels back to its reference, with units joules/coulomb or volt. This is naturally visualized by an electrical engineer as a capacitor storing charge. The charge is held immobile and the voltage across the capacitor is $v = C^{-1}q$, where $v$ is voltage, $q$ is charge and $C$ is capacitance. Capacitance $C$ is the phenomenological constant with the unit of farad (F) that translates charge to voltage. This is shown in Figure 1(a).

Consider now a collection of charges, flowing through a resistor. This is depicted in Figure 1(b). The rate of charge is current. The governing equation is $v = R \dot{q}$, which is the familiar $v = i R$ where $i = \dot{q}$. The over-dot in $\dot{q}$ represents derivative with respect to (w.r.t) time. The phenomenological constant is resistance $R$ with the unit ohm ($\Omega$).

One step further, the governing equation for the element that can respond to a rate of rate of charge, producing a resulting voltage is $v = L\ddot{q}$. The phenomenological constant is inductance $L$ with the unit of henry (H). We are more used to the standard form $v = L\frac{di}{dt}$ [2,3]. We show this in Figure 1(c).

It is possible to render the above information as a sketch in the charge-voltage domain. The sketch is often referred to as the chart or periodic table of fundamental elements. We have borrowed the expression "periodic table" as attributed to Chua[4].

## The periodic table of fundamental passive elements

|   | Z | Y | X | W |   |
|---|---|---|---|---|---|
| D |   | $U\dddot{q} = v$ | $L$ | $R$ | $\dddot{q}$ |
| C |   | $L\ddot{q} = v$ | $r\ddot{q} = \dot{v}$ | $C$ | $\ddot{q}$ |
| B | $L\dot{q} = \phi$ | $R\dot{q} = v$ | $C^{-1}\dot{q} = \dot{v}$ |   | $\dot{q}$ |
| A | $Rq = \phi$ | $C^{-1}q = v$ |   |   | $q$ |
|   | $\phi = \int v\, dt$ | $v = \dot{\phi}$ | $\dot{v}$ | $\ddot{v}$ |   |

*Figure 2: The periodic table of fundamental elements in the charge-voltage domain.*

The periodic table in Figure 2 has rows and columns of the grid labeled in upper case alphabets along the left and top edges. We will address a grid by its (row, column) label. The actual electrical variable that applies to each row or column is shown along the right and bottom edge. On the horizontal axis each column to the right is a time derivative of the one on its left. For example the x axis of (A, Y) is $\dot{\phi}$ which is the derivative of $\phi$, the x axis of (A, Z). Similarly along the y axis, each higher row's y axis is the derivative of the one below it.

Existing fundamental elements satisfy the following rules. We expect the same compliance from the memristor.

(i) Rule 1: Only one fundamental element can occupy a slot in the periodic table.
(ii) Rule 2: A transient event will not count as satisfying a constitutive relationship.

Rule 1 takes guidance from the periodic table of chemical elements which organizes elements based on atomic number. The periodic table of electrical elements organizes its elements based on the n[th] derivative of charge that relates the phenomenological constant to the voltage developed across the device.

Rule 2 is inferred from the definition of existing fundamental elements namely $C$, $R$ and $L$, where the phenomenological constant is linear.

## Locating the fundamental elements

The known fundamental elements from the preceding discussion appear along column Y rows A, B and C. This placement was done by comparing the governing equation of the slots to the governing equations in our earlier discussion. Moving into column X, slot (A, X) is empty because there is no known element that satisfies its rule. Slot (B, X) satisfies the derivative based definition of a capacitor namely $C \frac{dv}{dt} = i$ or the form that maps to our periodic table $C^{-1} \dot{q} = \dot{v}$. Similarly in column X, we observe that (C, X) is $r \ddot{q} = \dot{v}$ which is the small signal definition of a resistor, equivalent to $r = \frac{dv}{di}$. Inspection shows that each of the known fundamental elements travels along diagonals in our periodic table, leading to the n[th] derivative representation in terms of charge and voltage.

The thick vertical separator between columns Z and Y reinforces the idea that fundamental elements do not percolate into column Z. Columns to the right of Y contain derivatives of the governing equations from Y and do not constitute a fundamental definition because they are just mathematical operations. If there is place for a new fundamental element, it would be slot (D, Y) with a governing relation $U \dddot{q} = v$, equivalent to $v = U \frac{d^2 i}{dt^2}$, where $U$ is some yet to be discovered phenomenological constant. However, we will demonstrate later that an occupant of (D, Y) will be active, hence neither passive nor fundamental.

Before discussing column Z, let us review the governing equation that relates voltage to magnetic flux; namely Faraday's law. Faraday's law states that the negative of the rate of change of magnetic flux ($\phi_B$) will be equal to the electric potential ($\epsilon$) developed in the element, such as an inductor.

$$\epsilon = -\frac{d\phi_B}{dt} \qquad (1)$$

Assume that the experimenter is forcing a change in magnetic flux. The negative sign implies that the resulting voltage will be in such a direction as to generate a current whose magnetic field will try to oppose the forcibly induced change in flux. The equation is intended for use in a situation where the flux is truly a magnetic flux. However we notice that we could integrate the left hand side of equation (1) to result in $\phi = -\int \epsilon \, dt$ without insisting on a magnetic field. We have used $\phi$ without the subscript to denote the computed flux rather than the real magnetic flux. While this approach is mathematically correct, it gives rise to a possible misuse of the term flux. Given that we already have a charge-voltage plane to inspect, let us accommodate the definition $\phi = -\int \epsilon \, dt$ and disregard the need for a magnetic flux.

This moves our discussion into column Z in Figure 2. All elements along column Z generate flux $\phi = \int v \, dt$. Here $v \equiv \epsilon$ and represents voltage (electric potential). The missing negative sign (w.r.t a magnetic context) only serves to dictate the direction of voltage and there is no loss of accuracy for the discussion by leaving it out. Of the three known fundamental elements, only the inductor and resistor appear in column Z. As for the capacitor, we could mathematically write $\phi = C^{-1} \int q \, dt$ and be correct. However this would put us in a *meaningless* slot below (A, Z). In this system with passive elements, the capacitor did not make it into column Z because charge on a capacitor cannot be meaningfully transformed into the integral of voltage. This example with the capacitor shows how to create an infinite grid periodic table.

Turning our attention to slot (A, Z), the generalized governing equation is $U\, q = \emptyset$. We have temporarily introduced the phenomenological constant $U$ to stand in for what we might discover. Simple transposition gives $U = \frac{\emptyset}{q}$. We can rewrite this based on its time derivative form where the over-dot represents derivative w.r.t time.

$$U = \frac{\dot{\emptyset}}{\dot{q}} = \frac{d\phi}{dq} = \frac{v}{i} \tag{2}$$

Equation (2) defines the resistor, making $U$ in (A, Z) equal $R$ as shown in Figure 2. Prodromakis et al. among others use the intermediate expression $\left(\frac{\frac{d\emptyset}{dt}}{\frac{dq}{dt}}\right)$ containing the reference to charge, to suggest that $U = M = M(q(t))$ [5]. There are two issues here.

(i) The equation violates Rule 1 because the simple resistor will also suitably occupy slot (A, Z).
(ii) Any phenomenological constant inferred from time dependent intermediate equations is tenuous and violates Rule 2.

# Comparison with Strukov's table of fundamental elements

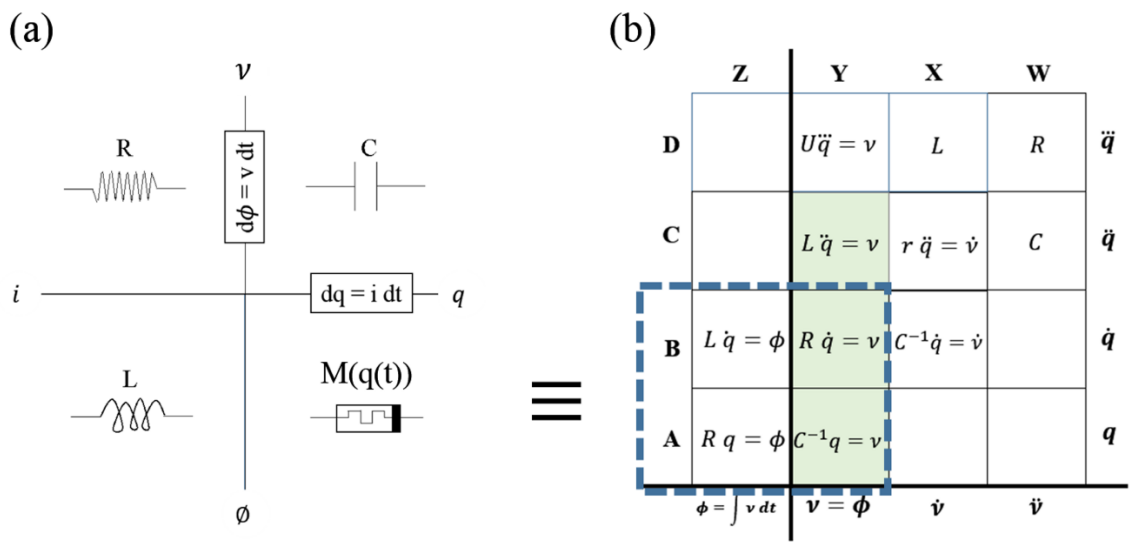

*Figure 3: Strukov's chart of fundamental elements compared to our representation. (a) Strukov's chart transcribed. (b) Our representation showing Strukov's chart as a subset.*

Figure 3 compares the periodic table of fundamental elements published by Strukov et al. at Hewlett Packard (HP) with that proposed by this paper[6]. We see that a left ninety degree rotation on the grid (A, Z)-(B, Y) in Figure 3(b) definitely makes it resemble Strukov's chart. There is a mistake in that original chart of fundamental elements that has carried over into many papers[6–8]. The framed expression on the positive x axis, $dq = i\, dt$ conflicts with the positive x axis being simultaneously labeled $q$. Algebraic manipulation of the framed expression gives $\frac{dq}{dt} = i$; implying that current equals the charge label along the positive x axis. This mistake is almost always overlooked. Nalawade et al. have corrected this but manage to label the flux axis incorrectly[9]. Kvatinsky et al. simply retain the empty frames[10]. Kumar has correctly labeled the chart's axes in an IETE technical review[11].

From the side by side comparison in Figure 3, we observe that the lower part of Strukov's chart maps to our column Z in Figure 3(b). Occupants of column Z are not fundamental because their relationships are defined by mathematical integration, requiring initial conditions. Even if we allowed such elements into the fundamental fold, the position proposed for the memristor is already occupied by the resistor.

The charge-voltage plane is sometimes referred to as the charge-flux domain, which is just the same as Figure 2, including the forbidden column Z. The memory resistor of 2008 with its phenomenological constant $M$, does not find a place in the charge-voltage plane because of the following two reasons.

(i) Slot (A, Z) is already occupied by the resistor. Rule 1 prohibits a second occupant.
(ii) A memristor with $M(q(t)) = \frac{\phi}{q}$ can only be evaluated by integration, violating Rule 2 by requiring a time-window for the integral.

# The electric-magnetic periodic table of passive elements

A moving charge generates a magnetic field. Ampere's law can be used to deduce that a current carrying wire will produce magnetic field lines perpendicular to the wire, in the direction suggested by the right hand rule. This means that Figure 1(b) and (c) have magnetic contributions as marked by the dashed curved arrows. This motivates us to expand the periodic table of fundamental elements into the magnetic plane.

| | $\widehat{W}$ | $\widehat{X}$ | $\widehat{Y}$ | $\widehat{Z}$ | Z | Y | X | W | |
|---|---|---|---|---|---|---|---|---|---|
| D | R | L | | | | | L | R | $\dddot{q}$ |
| C | C | $r\,\ddot{q} = -\ddot{\phi}_B$ | $L\,\dddot{q} = -\dot{\phi}_B$ | | | $L\,\ddot{q} = v$ | $r\,\ddot{q} = \dot{v}$ | C | $\ddot{q}$ |
| B | | $C^{-1}\dot{q} = -\ddot{\phi}_B$ | $R\,\dot{q} = -\dot{\phi}_B$ | $L\,\dot{q} = -\phi_B$ | $L\,\dot{q} = \phi$ | $R\,\dot{q} = v$ | $C^{-1}\dot{q} = \dot{v}$ | | $\dot{q}$ |
| A | | | $C^{-1}q = -\dot{\phi}_B$ | $R\,q = -\phi_B$ | $R\,q = \phi$ | $C^{-1}q = v$ | | | $q$ |
| | $-\dddot{\phi}_B$ | $-\ddot{\phi}_B$ | $v = -\dot{\phi}_B$ | $-\phi_B$ | $\phi$ | $v = \dot{\phi}$ | $\dot{v}$ | $\ddot{v}$ | |

*Figure 4: Locating fundamental elements in the electric-magnetic plane.*

Figure 4 is the extended version of Figure 2, with the charge-voltage plane to the right of the dashed blue center spine and the charge-flux (magnetic) plane to the left of the dashed blue center spine. The flux on the magnetic side is the true magnetic flux represented by $\phi_B$. Charge and its derivatives are along the y axis. The columns are labeled with hats on the magnetic side; row designators are shared among both domains. Discussion will address slots in (row, column) style. All devices within columns Z and $\widehat{Z}$ needed initial conditions to be specified to their derivative forms from Y and $\widehat{Y}$. To the left of center, each column leads into the n[th] derivative of the true magnetic flux. To the right, each column similarly leads to the n[th] derivative of computed flux, defined as the integral of voltage w.r.t time. Fundamental elements are in light green boxes that represent their constitutive relations. We start the discussion with the well-known candidates.

The capacitor from the charge-voltage plane does not make an appearance on the magnetic side because there is no magnetism involved for stationary charges. This automatically eliminates all the magnetic diagonals with capacitance $C$, like $(A, \widehat{Y})$, $(B, \widehat{X})$, $(C, \widehat{W})$ and so on shown in red dotted boxes.

With respect to inductors, we notice immediately that the inductor exists as it should in slot $(C, \widehat{Y})$. The inductor continues into slot $(B, \widehat{Z})$, crosses into the electric domain at $(B, Z)$ and further into slot $(C, Y)$. The inductor can live in both planes. An inductor generates a voltage proportional to the rate of change of current.

$$v = L\,\ddot{q} \tag{3}$$

The inductor also creates a true magnetic field around the coils for alternating current conditions in $(C, \widehat{Y})$. The slot $(C, \widehat{Y})$ is green because it is a fundamental relation in the magnetic domain.

$$-\dot{\phi}_B = L\,\ddot{q} \tag{4}$$

Under direct current conditions, the inductor satisfies the following relationship in $(B, \widehat{Z})$ and $(B, Z)$.

$$-\phi_B = L\,i = \emptyset \tag{5}$$

We recognize that $(B, \widehat{Z})$ is a one way relationship where a constant current can produce a constant magnetic field but not vice versa. Similarly $L\,i = \emptyset$ in $(B, Z)$ requires integration. Therefore equation (5) is not the constitutive relation for an inductor. Equation (4) from slot $(C, \widehat{Y})$ or equation (3) from $(C, Y)$ is the true constitutive relation for the inductor because it alone describes the ability of the inductor to bridge the magnetic and electric domains[2,3].

Let us now presume that the original postulate about a magnetic memristor is true. This will mean that the device should occupy slot $(A, \widehat{Z})$ based on the constitutive relation $M = -\frac{\phi_B}{q}$, with proper sign and subscripting. At this point we don't have enough data to contest this existence. We know from the loci of fundamental elements, that an incumbent in $(A, \widehat{Z})$ must also live in $(B, \widehat{Y})$. The governing equation for slot $(B, \widehat{Y})$ is $U\,\dot{q} = -\dot{\phi}_B$. This rule requires that there is some device with phenomenological constant $U$ which will produce a magnetic field that changes at a constant rate when a constant current $(\dot{q})$ is passed through it. Consider the 2008 HP (or 1971 Chua's) memristor for this slot. Pushing a constant current into the memristor will cause the voltage across the device to change $(\dot{v})$ as its resistance changes. The constant stimulus current will however only create a constant magnetic field, definitely not $-\dot{\phi}_B$. Therefore the 2008 HP (and 1971 Chua's) device cannot occupy $(B, \widehat{Y})$ except for the trivial condition $-\dot{\phi}_B = 0$. A resistor also satisfies this trivial condition. Let us try to derive the rule for the observed $\dot{v}$ from basic laws. Differentiating Faraday's law w.r.t time gives $\frac{d}{dt}\epsilon = -\frac{d}{dt}\left(\frac{d\phi_B}{dt}\right)$, which is $\dot{v} = -\ddot{\phi}_B$. We see $-\ddot{\phi}_B$ in slot $(C, \widehat{X})$. This requires a device to generate a rate of rate of magnetic flux when a rate of current passed through it i.e. $r\frac{di}{dt} = -\ddot{\phi}_B = \frac{dv}{dt}$. In other words, pushing a rate of current into the device should result in a rate of change of voltage $\dot{v}$; which will occur across the 2008 HP and 1971 Chua's devices. However, the change of voltage is only momentarily nontrivial and will evaluate to zero as soon as the transition from low to high resistance (or vice versa) is over; making this a transient event and violating Rule 2. Therefore there is no place in $(C, \hat{X})$ for the 2008 (or 1971) memristor. By inference, if $(C, \hat{X})$ and $(B, \hat{Y})$ are devoid of memristors, then $(A, \hat{Z})$ is also forbidden to the memristor. Additionally, there is no incumbent possible in $(A, \hat{Z})$ because stationary electric charges cannot produce a magnetic field. Thus we have colored all the excluded squares of the periodic table in red, eliminating the 2008 HP (and 1971) device from the magnetic plane. Proposing that $q$ in $(A, \hat{Z})$ is the integral of current and not a literal stationary charge, puts us in the realm of abstraction-by-integration. Then the memristor is an abstract device that responds to an abstract electric charge. In any case, a resistor already satisfies the conditions of $(A, \hat{Z})$ even in this realm of abstraction.

Can the 1971 postulate be probable *anywhere* in the magnetic side? The loci of fundamental elements suggests that nothing other than a form of resistor can occupy the trajectory $(A, \hat{Z})$, $(B, \hat{Y})$, $(C, \hat{X})$, $(D, \hat{W})$ etc. proposed for the memristor. The memristor in any manifestation is therefore excluded from the magnetic and electric side of the periodic table by Rule 1 and Rule 2.

## The linear-nonlinear debate

The notion that the memristor postulated in 1971 may somehow exist as a nonlinear yet fundamental entity can be dispelled by reviewing Chua's seminal paper. Section III of Chua's paper states that when the memristor $\phi$-$q$ curve is a straight line, the memristor reduces to a *linear time-invariant resistor*; very much in keeping with our findings in the prior sections[12].

Now consider the nonlinear case. Chua states that "…only memristors characterized by a monotonically increasing $\phi$-$q$ curve can exist in a device form without internal power supplies." A nonlinear $\phi$-$q$ curve will always have a positive, albeit variable slope. However, the ratio of $\phi$ to $q$ still has the units of ohm, *without a phase shift*, making this a nonlinear resistor. Simple pn diodes or transistors as shown in Table 1 can emulate nonlinear resistors. In the context of the three known fundamental devices, Tour et al. state that "The behavior of each of these elements is described by a simple linear relationship...", clearly affirming that linearity is central to being fundamental[7]. Nonlinear resistors are not fundamental because they can be modeled by an assembly of piecewise linear components.

The report from Di Ventra et al. and general modeling knowledge shows that only an active element can produce a negative differential resistance (NDR) which is seen in the memristor current-voltage curves[13]. NDR eliminates the nonlinear passive diode from being able to model the memristor, leaving active circuits as the only option to model memristors.

Let us review the memristor in the light of Strukov's expression for the phenomenological constant $M(q(t)) = \frac{\phi}{q}$. When $M(q(t))$ is a positive constant and $\frac{d}{dq}M(q(t)) = 0$, then the device is a *linear time-invariant resistor* which is the already known fundamental element $R$. All other cases are at the very least nonlinear and excluded from being fundamental. By virtue of nonlinearity and *ignoring* any activity criterion, the memristor should occupy $(C, X)$ in Figure 4.

The suggestion that a circuit element can be a fundamental passive by virtue of nonlinearity is fallacious. If this were true, then the small signal resistor $r$ in $(C, X)$, which could represent a passive diode, would also be a fundamental element. Even if we choose to henceforth recognize nonlinear passive devices as fundamental in some "expanded design space" as suggested by Williams et al., (i) the memristor cannot occupy $(A, Z)$ or $(A, \hat{Z})$ because those slots are occupied by the linear resistor and (ii) the memristor cannot occupy $(C, X)$ because in the current-voltage domain the device exhibits hysteresis – an active phenomenon, thereby excluding it completely from any table of fundamental passive elements[6,12,14]. Slot $(C, \hat{X})$ was rejected in previous discussion.

## Dissecting the memristor

If not fundamental, what is it? The first evidence is the original model from HP. It looks like a potentiometer made of two resistors and a slider[6]. The slider must be moved as a function of time to make the device transition between the low and high resistance states. In spite of the many shortcomings of the HP model it captures the essence of the memristor – a two terminal series connection of resistors with a low resistance $R_{LO}$ and high resistance $R_{HI}$, exhibiting NDR. While Di Ventra et al. argue that a negative resistance can only ensue from an active element, the memristor community does not readily equate NDR with the presence of an active element[13]. Through the clever use of window functions that are arbitrarily introduced into equations, HP hides the presence of active elements in their memristor model. After all, a potentiometer has no inherent hysteresis.

Apart from very specific and focussed claims about activity and hysteresis, we seek a general model that can demonstrate these disqualifying qualities and memristive behavior. An original research in symbolic modeling has revealed two impedances that torsion in the complex plane[15,16]. That work proposes to model vacancy migration with the generic transport equation, starting with the conservative form; with vacancy concentration $u$, vacancy velocity $v$ and the accumulation (or shock wave) boundary $x_b(t)$. In the following discussion, $n_b(t) = x_b(t)/d$, implying normalization.

$$\left(\frac{\partial}{\partial t} + \frac{\partial}{\partial x} v\right) u = \frac{Du}{Dt} \tag{6}$$

The total derivative $Du/Dt$ in a conserved system is zero. Expanding the operators from (6),

$$u_t + (v - x_b'(t))u_x + u\, v_x = 0. \tag{7}$$

Apply the transformation $(v - x_b'(t)) \to \vartheta$ where $\vartheta$ is termed the local wave velocity. This results in $u_t + \vartheta\, u_x + u\left(\vartheta + x_b'(t)\right)_x = 0$. Guarantees from the Method of Characteristics (MoC) allow some simplifications. The first spatial derivative in the third term will evaluate as $\vartheta_x = 0$ because the MoC ensures that $\vartheta|_{x \to x_b(t)} = k$, $k$ constant. The second term $x_b'(t)_x = 0$ because $x_b(t)$ is independent of $x$. This immediately leads to the variable coefficient advection (VCA) equation, which is the starting point for the analytical solution. Equation (8) is also *very similar* to a Burgers' equation with a variable coefficient for $u_x$.

$$u_t + \vartheta\, u_x = 0 \tag{8}$$

The logistic function is presented as the solution to this variable coefficient (Burgers') equation.

$$u = \left(1 + a\, e^{-f_0 \emptyset (x - x_b(t))}\right)^{-1} \tag{9}$$

The published results for (9) include the identification of physical limiters and effect of doping on switching time, resistance range, emergence of ohmic/semi-conducting behavior as a function of temperature etc. When used in conjunction with arbitrary variables acting as tuning knobs, the estimations agree reasonably with independent empirical data[15,16]. These references also explain the complete list of variables involved in the computation.

Computation shows that the local wave velocity $\vartheta$ at the location of the moving accumulation boundary is a constant, except for the initial acceleration and final deceleration, as expected. The extreme initial acceleration from rest is omitted from Figure 5, showing only the slowdown when approaching device center (at the blue dot-dashed line) and the device ends (at the red dotted line).

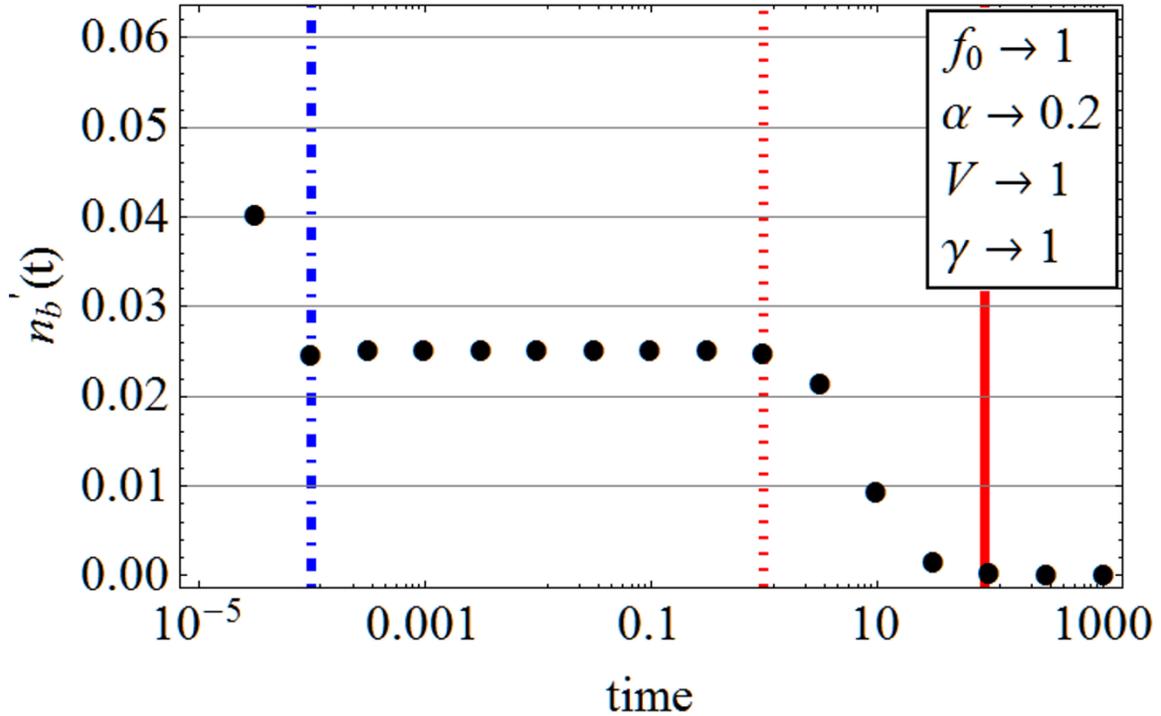

Figure 5: The normalized accumulation boundary velocity. The blue dot-dashed line indicates the entry of the shock wave into the device and the red dotted line indicates the start of slowdown.

$$\vartheta|_{x=x_b(t)} = -\frac{u_t}{u_x} = -x_b'(t) \tag{10}$$

The vacancy evolution profile in (9) can be transformed into resistance[15].The Burgers' model reveals the memristor as the sum of two impedances. The reactive components always sum to zero. Memristor resistance $R(\emptyset) = Z_1(\emptyset) + Z_2(\emptyset)$; with the integral of voltage as flux $\emptyset$.

$$Z_1(\emptyset) = -\frac{p \ln((\alpha-1)e^{f_0(n_b(\emptyset)-1)\emptyset} + (p-1)\alpha)}{f_0(p-1)\emptyset} - \frac{1}{2(p-1)} \quad (11)$$

$$Z_2(\emptyset) = \frac{p \ln((\alpha-1)e^{f_0 n_b(\emptyset)\emptyset} + (p-1)\alpha)}{f_0(p-1)\emptyset} - \frac{1}{2(p-1)} \quad (12)$$

The complex resistors evaluate as $Z_1 = \pm a \mp j\,b$ and $Z_2 = \mp c \pm j\,b$, where the positive term from among $a, c$ is always larger. Therefore the composite resistance is always positive. The model unambiguously reveals the presence of a non-dominant negative resistance. It is possible to associate the negative impedance (resistance, reactance) with the shockwave that Tang et al. have deduced[17]. The Burgers' model locates the shockwave at the accumulation boundary.

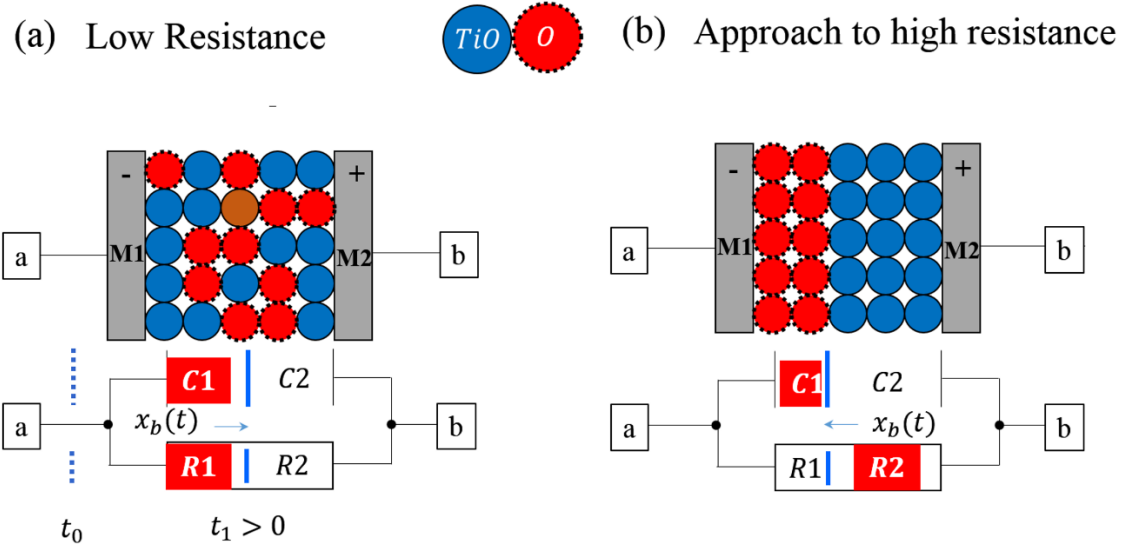

Figure 6: The traveling shockwave gives rise to negative (resistance, reactance) pairs. Negative components are shown in reversed white-on-red background.

The negative resistance is not visible to an external observer except during transition and is *indispensable* for representing flux dependent hysteresis which is key to memristor functionality.

Figure 6(b) shows the device as two capacitors in series; that combination in parallel with two series resistors. The normalized capacitive impedance can be modeled as $X_C = \frac{1}{j\omega\varepsilon_0\varepsilon_r A}(n_b(t)|1 - n_b(t))$; $n_b(t) = x_b(t)/d$. The static impedance across the memristor is always a resistance while memristance is an AC phenomenon[12]. We are therefore interested in the rate of change of impedance which precipitates a dynamic response, all variables constant except for $x_b(t)$.

$$\{\dot{X}_{C1}, \dot{X}_{C2}\} = \left\{-\frac{j}{\omega}n'_b(t), \frac{j}{\omega}n'_b(t)\right\} \quad (13)$$

All else remaining unchanged, the dynamic impedance is determined by the sign of the time derivative of $x_b(t)$. When the accumulation boundary (shockwave or reference point) is accelerating, $\dot{X}_{C1} < 0$ and $\dot{X}_{C2} > 0$ and vice versa when decelerating. These components disappear when the accumulation boundary is stationary.

A similar argument can be presented for the dynamic resistance. Defining resistance as $R = \frac{\rho}{A}(n_b(t)|1 - n_b(t))$ and differentiating w.r.t time we get the following.

$$\{\dot{R}1, \dot{R}2\} = \{\rho_1 n'_b(t), -\rho_2 n'_b(t)\} \tag{14}$$

Once again we see that the two component resistors have values that are functions of the resistivity. We naturally expect the resistance to electronic flow to be modulated by the distribution profile of the positively charged, mobile vacancies. Unlike dielectric constant, resistivity is a function of the vacancy concentration.

Without the hysteresis induced negative components precipitated by the shockwave, memristor current-voltage curves cannot exhibit lobes. The boundary between the low and high vacancy concentration regions in the memristor emulates the slider of a rheostat, partitions the device into two (series) resistors and implements (active) hysteresis. Hysteresis is acknowledged by many authors including Chua, Williams, Strukov and Biolek[6,12,14,18,19]. Hysteresis can be implemented in circuit with the operational amplifier, Schmitt trigger or voltage/current-controlled elements – all of which are active.

The Chua Memristor Center's website claims that Corinto et al. have constructed a memristor model with one port passive components[20,21]. This is ***impossible*** if the memristor is a fundamental element. Another peer reviewed publication associated with the circuit replaces the purported passive resistor with a Chua's diode; which is a locally active device[22]. Any activity implies a negative resistance[23]. Memristor modeling has always needed active elements because it is implicitly active[12,24,25].

The Chua Lectures, Part 3 demonstrates that the ideal memristor will draw infinite current $i(t) = 3v(\emptyset(0) + v\,t)^2|_{t\to\infty} \to \infty$ for any non-zero voltage stimulus. The square law relationship w.r.t time hints at an unphysical active element[26]. This is remarkable because none of the other three fundamental elements are unphysical. A device with $i(t) \propto t^2$ could in theory occupy (D, Y) in Figure 4. The "Locating the fundamental elements" section had suggested that (D, Y) *may* harbor a new fundamental device. Our mystery device could in theory be multiplying two inductor currents, each of $i(t) = L^{-1}v\,t$. Multiplication however requires active elements thereby eliminating passive devices from ever occupying (D, Y). Therefore the set of fundamental elements is strictly limited to $C$, $R$ and $L$ along ([A, B, C], Y) in Figure 4[27].

## Conclusion

It is incorrect to state that current, voltage, charge and flux-(linkage) are fundamental circuit variables and establish relationships among them. From our discussion it should be amply clear that voltage developed in response to electrical charge separated from its reference plane forms the foundation of electrical engineering. Fundamental devices are physical entities that translate the various states of motion of charge to a measurable voltage (or magnetic field) that arises in (or around) the device. The chart of fundamental elements from Strukov et al., transcribed in Figure 3(a) becomes untenable when we define and adhere to rules for the creation and population of the periodic table of fundamental elements. Neither the magnetism nor electric flux based memristor finds a place in the periodic table due to multiple violations.

Supporters promote that the memristor is fundamental because it cannot be modeled with the traditional $C$, $R$ and $L$[14] although contradicting themselves with a ***purported*** passive model[21]. We respond that memristors cannot be modeled with $C$, $R$ and $L$, not because they are fundamental but because memristors are nonlinear resistor composites that exhibit active hysteresis when switching from low to high resistance or vice versa.